# Statistics of Non-Rayleigh Speckles Generated in Nonlinear Media

*Deependra Singh Gaur [1], Akanksha Gautam [2], Rakesh Kumar Singh [2], and Akhilesh Kumar Mishra [1, 3, *]*

[1] *Department of Physics, Indian Institute of Technology Roorkee, Roorkee 247667, Uttarakhand, India*
[2] *Laboratory of Information Photonics & Optical Metrology, Department of Physics, Indian Institute of Technology (Banaras Hindu University), Varanasi 221005, Uttar Pradesh, India*
[3] *Centre for Photonics and Quantum Communication Technology, Indian Institute of Technology Roorkee, Roorkee 247667, Uttarakhand, India*
[*]akhilesh.mishra@ph.iitr.ac.in

**Abstract**

We analytically derive an expression for a speckle field's intensity probability density function (PDF) in a nonlinear medium. The analytically driven results are in good agreement with the numerical outcomes. In a focusing nonlinear medium, the local intensity of the speckle is enhanced as manifested through the longer tail of the PDF. In contrast, the local intensity of speckle is reduced in the presence of a defocusing nonlinearity, and the tail of the probability density function also reduces. This change in local intensity of the speckles arises due to the cubic Kerr nonlinearity, which eventually modifies the second-order statistics. Hence, the intensity correlation is altered as per the nature of the associated nonlinearity while the field correlation remains invariant of both types of the nonlinear conditions.

*Keywords:* Speckle, Kerr nonlinearity, correlation.

**Introduction**

The scattering of a coherent wave when transmitting through a random medium or reflecting from a rough surface inflicts a transformation in the complex amplitude and phase of the wave. Multiple scattered partial waves interfere with each other and form a granular structure known as a speckle pattern [1,2]. This phenomenon has been observed in a wide range of waves of different nature [3,4]. Speckle field is the superposition of a large number of partial waves with random and independent amplitudes and phases, where the phases are distributed over a range of $2\pi$. Under these general conditions, the speckle field is characterized by Rayleigh statistics where the field's amplitude follows a Gaussian distribution, and a negative exponential intensity distribution is observed in the speckle [1,2]. The central limit theorem of random variables dictates the Rayleigh statistics under these circumstances [5]. For some instances, such as in the weak scattering regime or near field region of the scattering media, where limited scattering waves are detected, random phases are not distributed uniformly in the range of $2\pi$. Therefore, statistical properties follow non-Rayleigh statistics such as the Rician or Rice distribution [1,5-8]. Statistical properties deviate from the Rayleigh distribution even in the strong scattering regime if the phase of partial waves is correlated. The mesoscopic correlations explain the existence of non-Rayleigh intensity statistics in a scattering medium [9-11].

Non-Rayleigh speckles have numerous potential applications in imaging and microscopy, but these speckles are either underdeveloped or partially coherent, which revert to Rayleigh speckles after a limited propagation range [12-15]. These applications demand fully developed, robust non-Rayleigh statistics that can be tailored. This requirement can be fulfilled by considering the interference of many partial waves with uniformly distributed phases with statistically dependent complex amplitudes [1,2]. Spatial light modulation (SLM) alters the intensity distribution of a speckle pattern by introducing higher-order correlations that result in super-Rayleigh and sub-Rayleigh speckles with enhanced and diminished contrast, respectively [16]. Further, SLM paves the way to customize the intensity statistics by using a phase-only method for tailoring the desired intensity statistics, and the intensity transformation statistics can be described analytically [17-21]. Notably, transverse spreading of the speckle pattern under the effect of diffraction affects the volumetric imaging applications. However, nondiffracting speckles with non-Rayleigh intensity statistics are reportedly generated using SLMs [22,23].

Rayleigh speckles obey the Siegert criterion, which is violated by non-Rayleigh speckles. Light scattering in a nonlinear medium also generates speckles with non-Rayleigh intensity statistics, affecting interference results. Intensity-dependent refractive index alters the interaction among photons, which can be attractive or repulsive, depending upon the focusing or defocusing nature of the cubic Kerr nonlinearity [24, 26]. This observation opened a new avenue to realize speckles in a nonlinear medium with tailored intensity statistics that can be employed in different potential applications [27-29]. The available literature reports on the experimental and numerical studies of the of non-Rayleigh speckles describing intensity correlation and PDF in nonlinear media. However, an analytical relation depicting the dependence of PDF on nonlinear index change and a discussion on first order statistics in nonlinear media are still lacking.

In the present work, we show that light scattering in a nonlinear medium generates non-Rayleigh speckle pattern. Intensity distribution of non-Rayleigh speckle deviates from the typical negative exponential behaviour observed in linear media. In the following, we derive an analytical expression of the intensity PDF in nonlinear media, which predicts the intensity distribution in both linear and cubic Kerr nonlinear media. A consideration of the first-order statistics is also undertaken. The field distribution of the generated speckle pattern follows Gaussian statistics and remains invariant in all nonlinear conditions, while the intensity statistics are altered with the medium nonlinearity. Focusing nonlinearity generates an intense speckle pattern that leads to a longer tail of the intensity PDF, resulting in non-Rayleigh speckles. On the other hand, defocusing nonlinearity enhances the divergence of speckles, which leads to a reduction in the PDF tail.

**Theory**

In this section, we present a theoretical model that predicts the intensity distribution of a speckle field in a nonlinear medium. Consider a speckle field that can be described as $\vec{E} = |E|\exp(i\phi)$ and their intensity $I = |\vec{E}|^2$. This relation between field and intensity holds only in a linear medium where the strength of the field is too weak to polarize the medium. On the other hand, strong electric field interaction induces polarization in the medium, and the nonlinear response of the medium starts to play a significant role. In this case, the susceptibility of the medium can be Taylor expanded, and the refractive index of the medium can be modified. Therefore, the intensity in such a medium can be represented as [30,31]

$$I_{NL} = |\vec{E}|^2 + n_2|\vec{E}|^4 = I + n_2 I^2 \quad . \tag{1}$$

$$I_{NL} = I + n_2 I^2 \tag{2}$$

$$n_2 I^2 + I - I_{NL} = 0. \tag{3}$$

$$I = \frac{-1 \pm \sqrt{1 + 4n_2 I_{NL}}}{2n_2}. \tag{4}$$

The above expression relates to the intensity $I_{NL}$ in a nonlinear medium to the intensity $I$ in a linear medium. Here, we use the variable transformation method to derive the PDF of the intensity in nonlinear media. According to this transformation, if two random variables $I$ and $I_{NL}$ are connected through the monotonic transformation such that $I_{NL} = f(I)$ then the PDF of both random variables is related through [1,5]

$$\rho_{I_{NL}}(I_{NL}) = \rho[f^{-1}(I_{NL})] \left| \frac{dI}{dI_{NL}} \right|. \tag{5}$$

where $\rho$ and $\rho_{I_{NL}}$ are the PDF in linear and nonlinear media respectively and $f^{-1}$ is inverse of $f(I)$. The PDF of intensity in linear medium obeys the Rayleigh distribution and can be expressed as $\rho(I) = \frac{1}{2\sigma^2} \exp\left(-\frac{I}{2\sigma^2}\right)$ where $\sigma$ is the standard deviation in intensity from the mean value. Therefore, the new distribution in nonlinear media can be expressed as

$$\rho(I_{NL}) = \frac{1}{2\sigma^2 \sqrt{1 + 4n_2 I_{Nl}}} \exp\left[\frac{1 \mp \sqrt{1 + 4n_2 I_{NL}}}{4n_2 \sigma^2}\right] \tag{6}$$

Here, only the negative sign has physical meaning; therefore, we omit the positive sign from the exponent in equation (6).

To validate our model, we study the propagation of a 2D speckle field under the paraxial approximation. The numerical simulation to model the propagation employs the nonlinear Schrödinger equation (NLSE). The dimensionless NLSE can be expressed as [24,26]

$$\frac{\partial E}{\partial z} = \frac{1}{2}\left(\frac{\partial^2 E}{\partial x^2} + \frac{\partial^2 E}{\partial y^2}\right) + n_2 |E|^2 E, \tag{7}$$

where $z$ is the normalized propagation coordinate and $x$, $y$ are the normalized transverse coordinates scaled with arbitrary length. The nonlinear refractive index change is associated with the normalized $n_2$. The normalized speckle field is denoted with $E$.

**Results and Discussion**

We numerically generate a speckle field and study the propagation dynamics in linear and nonlinear media, assuming identical scattering conditions. In our simulation, we consider a laser beam of wavelength $532\ nm$ with $2\ mm$ waist size that is illuminated on apertures of different sizes and shapes (see supplementary [32]). Beam coming out from a square aperture of side length $2\ mm$ propagates in the scattering medium_undergoes multiple scattering and the interference of the scattered waves lead to the formation of random intensity pattern. The intensity pattern at $z = 5$ is shown in fig. 1.

Fig. 1 (a) shows the fully developed speckles in a linear medium where $n_2 = 0$. In the presence of focusing nonlinearity ($n_2 > 0$), the nonlinear phase shift suppresses diffraction, and the self-focusing of speckles is observed, which results in enhanced local intensity of the speckle as depicted in Fig. 1 (b). On the other hand, defocusing nonlinearity ($n_2 < 0$) escalate the diffraction, and therefore, speckles are flattened, and local intensity is reduced, as shown in Fig. 1 (c) (see the colour bars).

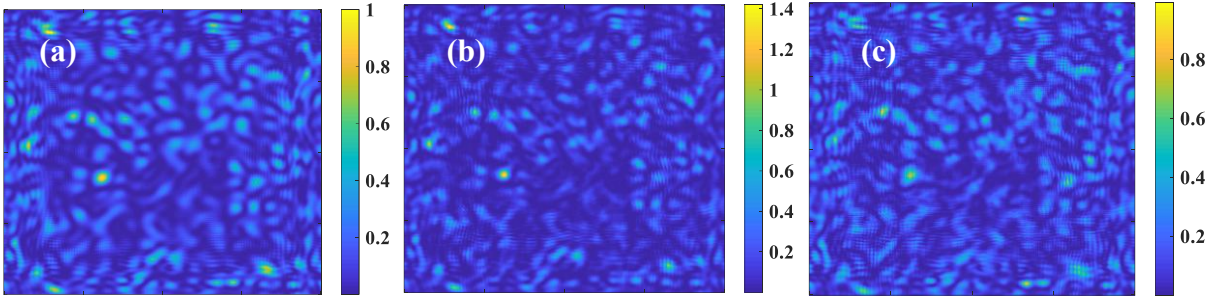

Fig. 1 The speckle intensity pattern for (a) linear $n_2 = 0$, and nonlinear (b) $n_2 > 0$ & (c) $n_2 < 0$ media.

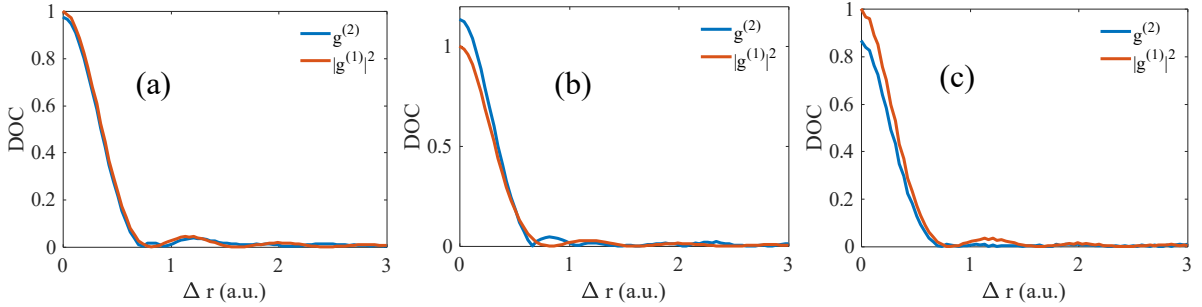

Fig. 2 Comparison of the intensity correlation function and square of the field correlation in (a) linear, (b) focusing, and (c) defocusing nonlinear media.

Further, we calculate the spatial correlation function of the speckle field and spatial intensity to reveal the impact of nonlinearity on the statistical properties of the generated speckle patterns. Speckle field correlation function $(g^{(1)})$ is given by $C_E(\Delta r) = <E(r)E^*(r + \Delta r)>$ while the intensity correlation function $(g^{(2)})$ $C_I(\Delta r) = \sum_m^M (I_1^m(0) - \overline{I_1(0)})(I_2^m(\Delta r) - \overline{I_2(\Delta r)}))/M$, where $M$ denotes the number of realizations [19, 33]. The normalized correlation function is called the degree of coherence (DOC), which varies with the distance.

The correlation function is evaluated by averaging the two-point correlation over 5000 realizations of the random patterns. It is well known that speckles generated in a linear medium follow the Siegert criterion; therefore, the square of the field correlation must be equal to the intensity correlation, as shown in Fig. 2 (a). Our numerical calculation confirms that the intensity distribution necessarily follows the Rayleigh statistics. But this relation is violated in nonlinear media as shown in Fig. 2 (b) and 2 (c). When $n_2 > 0$, intensity correlation is found to be larger than the square of the field correlation, as depicted in Fig. 2 (b). Moreover, the width of the intensity correlation function is also reduced. On the other hand, intensity correlation becomes smaller than the square of the field correlation in the presence of $n_2 < 0$ while its width is increased as shown in Fig. 2 (c).

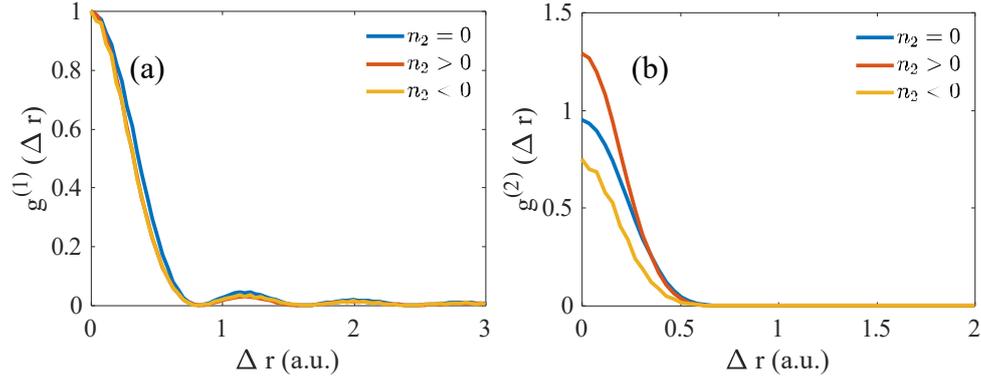

Fig. 3 (a) First-order correlation function, (b) second-order correlation for linear, focusing, and defocusing nonlinear media.

Fig. 3 (a) shows that the field correlation function remains the same for speckles generated in linear and nonlinear media. Nonlinearity-induced interaction does not influence the field statistics of the speckle pattern, while nonlinear effects modify the intensity correlation function. Focusing nonlinearity ($n_2 > 0$) leads to the self-focusing of the speckles, thus the local intensity of the speckle is enhanced. Hence, the peak value of the intensity correlation function increases correspondingly. On the contrary, defocusing nonlinearity ($n_2 < 0$) reduces the peak value of the intensity correlation function, as shown in Fig. 3 (b). To validate our numerical result, we further calculate the correlation functions for different apertures of different shapes and sizes, including square and circular, as detailed in the supplementary. Our results are found to be consistent in each measurement.

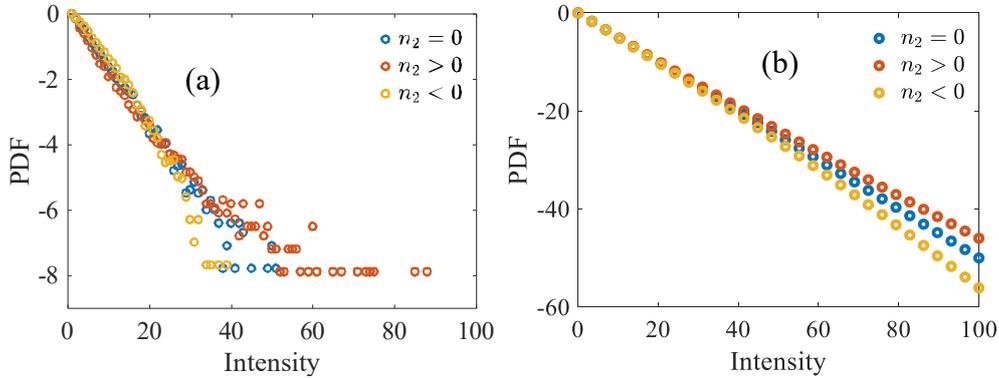

Fig. 4 (a) Numerically calculated probability density function and (b) probability density function from analytical model

The influence of nonlinearity can be further explored by examining the intensity probability distribution in both linear and nonlinear media. In case of $n_2 > 0$, speckles undergo self-focusing and high intensities are found probable owing to the enhanced local intensity. Therefore, the longer tail in the probability distribution appears as shown by the red circles in Fig. 4 (a). While for $n_2 < 0$, bright spots spread over the region and dark spots are diminished and hence high-intensity speckles no longer sustain. This eventually modifies the intensity probability distribution. Thus, the probability of high intensity values is relatively weak for $n_2 < 0$ and therefore tail of the probability distribution function is reduced as shown by the yellow circles in Fig. 4 (a). When $n_2 = 0$ (linear medium), the probability distribution falls between the above-discussed two cases as depicted in Fig. 4 (a). These numerical outcomes were best fitted by our analytical expression derived in Eqn. (6). The plots of the analytically derived intensity probability distributions are shown in Fig. 4 (b), which manifest the impact of nonlinearity-induced change in the intensity distribution. Fig. 4 (b) shows that the intensity tail is longer for $n_2 > 0$ while it reduces for $n_2 < 0$. In case of defocusing nonlinearity with $|4n_2 I_{NL}| > 1$, exponential part becomes imaginary, and our analytical expression fails to predict the PDF in the presence of defocusing nonlinearity. However, it can predict the PDF by focusing on nonlinearity.

**Conclusion**

The statistical properties of speckle patterns generated in linear and nonlinear media have been studied. The study reported that the field correlation function remains invariant while the intensity correlation depends upon the nonlinearity of the medium. In focusing nonlinearity, a longer tail in the intensity PDF was observed therefore correlation function was enhanced compared to the linear case. While the tail of the intensity PDF was reduced in the presence of defocusing nonlinearity, the speckle contrast was reduced, and the correlation function was decreased compared to that in the linear case. This change in PDF is mediated through the intensity-induced nonlinear refractive index, which modifies the local intensity distribution of the speckle pattern. Non-Rayleigh nature of the numerically calculated intensity PDF can be predicted by our theoretically derived expression. These findings provide the fundamental insights into the light matter interaction dynamics in complex nonlinear media.

**Acknowledgement**

Deependra Singh Gaur acknowledges to the Ministry of Education (MoE) India and IIT Roorkee for providing the fellowship assistance. Authors are also grateful to the Prof. Yaron Bromberg, Racah Institute of Physics, The Hebrew University of Jerusalem, Israel for helpful discussions.